\documentclass[12pt]{article}

\newcommand{\be}{\begin{equation}}
\newcommand{\ee}{\end{equation}}
\newcommand{\bi}[1]{\vspace{-3mm} \bibitem{#1}}
\usepackage{epsfig,amsmath,amssymb,graphics,graphicx}

\begin{document}


\begin{center}
{\Large \bf Fractional Dynamics of Systems 
with Long-Range Space Interaction and Temporal Memory}
\vskip 5 mm

{\large \bf Vasily E. Tarasov$^{1,2}$ George M. Zaslavsky$^{1,3}$ }\\

\vskip 3mm

{\it $1)$ Courant Institute of Mathematical Sciences, New York University \\
251 Mercer St., New York, NY 10012, USA }\\ 
{\it $2)$ Skobeltsyn Institute of Nuclear Physics, \\
Moscow State University, Moscow 119992, Russia } \\
{\it $3)$ Department of Physics, New York University, \\
2-4 Washington Place, New York, NY 10003, USA } 
\end{center}

\vskip 3mm

\begin{abstract}
Field equations with time and coordinates derivatives of noninteger order 
are derived from stationary action principle for
the cases of power-law memory function 
and long-range interaction in systems.
The method is applied to obtain a fractional generalization of 
the Ginzburg-Landau and nonlinear Schr$\ddot{o}$dinger equations.
As another example, dynamical equations for particles 
chain with power-law interaction and memory are considered in 
the continuous limit.
The obtained fractional equations can be applied 
to complex media with/without random parameters or processes.
\end{abstract}

\vskip 3mm

\section{Introduction}

From the contemporary vision of complex media, where microscopic processes take place 
and where many important applications are utilized, 
it is too far from considering the media as uniform gases, liquids, or solids. 
The most typical features of the new physical objects and/or processes are 
fractality of their structure and of intrinsic dynamics or kinetics. 
Observation of fractality of the basic processes 
began fairly long ago (see for review \cite{Montr,SZK}). 
Typically the complexity of systems is linked 
to the long term memory, long-range interactions, non-markovianity 
of the kinetics, and particularly with the Levy-type processes 
(Levy flights) \cite{Le}. 
The literature on this subject is vast. 
Let us mention some of the most related references, 
where the indication of the complexity can lead, in one or another way,  
to the fractional description of the dynamic and/or kinetic processes 
with fractional time \cite{MV,S1,Sh+SB}; 
systems of many coupled elements \cite{Ku,Win}; 
colloidal aggregates and chemical reaction medium \cite{PA83,SMWC}; 
wave processes \cite{ZL,MMT,Wy}; porous media \cite{Nig}; 
quantum mechanics and quantum field theory \cite{Berry,Laskin,Gf}; 
plasma physics \cite{Caa,Br6,San}; 
magnetosphere \cite{Za}; 
random processes and random walks \cite{Uch,Meersh+,SZ,MK,HMS}; 
fractional diffusion and Brownian motion \cite{Ta,YPP,Sh+SB}; 
weak and strong turbulence \cite{MMT,F,SWK+S-89}; 
fractional kinetics and chaos theory \cite{Zaslavsky7} 
(see for review \cite{Zaslavsky1,Zaslavsky2}).

It seems that the basic formal tool to be applied to is the description of 
the processes by fractional equations, i.e., by the ones 
that contain fractional derivatives or integrals \cite{SKM,OS,Podlubny,KST}. 
The theory of derivatives of non-integer order goes back 
to Leibnitz, Liouville, Riemann, Grunwald, and Letnikov \cite{SKM,OS}. 
Derivatives and integrals of fractional order have found many
applications in recent studies in physics because of their continually 
growing numerous applications. 

Usually, onset of fractional derivatives (integral) is linked to
different power type asymptotic interactions or time memories. 
Depending on what kind of specific features characterize the physical object, 
the fractional derivative (integral) can be with respect to time or space coordinate. 
In the description of particles transport, when the dynamics is chaotic, 
the fractional  derivatives emerge in space and time simultaneously 
as a natural reflection of scaling properties of the phase space dynamics 
\cite{Zaslavsky7,Zaslavsky1}. 
The diffusion described by the fractional equations is called the anomalous one. 
The occurrence of such derivatives could also be related 
to the space-time decay \cite{MGo,ZES}, i.e., 
to pure dynamical processes without kinetics or diffusion. 
Particularly it was shown in \cite{LZ,TZ3,KZT,KZ,JMP} how the long-range interaction 
between different oscillators can be described by the fractional differential equations 
in the continuous medium limit. Another way to connect the fractional equations 
with specific dispersion laws of the media was considered in \cite{MMT,ZL,Zaslavsky6}.

The goal of this paper is to provide a systematic approach to the onset of 
fractional equations as a result of existence of long-range interaction 
in a corresponding space and long-range time memory in the system of fields or 
particles depending on what kind of physical objects are considered. 
The notions of long terms memory or interaction can be exactly specified by 
power laws in time for a memory function and power law interaction between 
different elements of the medium. It is of importance to understand the conditions 
when the fractional derivatives (integrals) occur since it allows us to involve 
into the consideration power tools of fractional calculus. 

In Sec. 2, we consider the variation of action functional that 
describes field with memory and long-range interaction.
The long-time memory and long-range interaction can be introduced 
through power-like kernels of the action functional.
The corresponding powers are defined by the exponent $\alpha$
(for space) and $\beta$ (for time), which in general can be fractional.
The Euler-Lagrange equations lead to the equation
with fractional $(\alpha,\beta)$-derivatives.
In Sec. 3, the obtained results are used
for derivation of $(\alpha,\beta)$-generalization of the
Ginzburg-Landau and nonlinear Schr$\ddot{o}$dinger equations.
In Sec. 4, we consider chains of particles with 
long-range interaction and memory function.
Applying the results of Sec. 2, we derive the continuous limit 
of the particle dynamics equations.
In two Appendices, we provide a brief information on the 
Riemann-Liouville, Caputo and Riesz fractional derivatives
used in paper, and $n$-dimensional generalization of 
the final fractional equations.

\section{Action functional and its variation}

\subsection{Action functional}

Let us define the action functional as
\be \label{1} S[u]=\int_R d^{2} x \int_R d^{2} y \left( 
\frac{1}{2} \partial_{t} u(x) g_0 (x,y)  \partial_{t'} u(y) 
+\frac{1}{2} \partial_{r} u(x) g_1 (x,y)  \partial_{r'} u(y) 
-V(u(x),u(y)) \right) . \ee
Here $x=(t,r)$, $t$ is time, $r$ is coordinate, and $y=(t',r')$.
The integration is carried out over a region $R$ of the 
$2$-dimensional space $\mathbb{R}^2$ to which $x$ belong.
The field $u(x)$ is defined in a $2$-dimensional region $R$ of $\mathbb{R}^{2}$.
We assume that $u(x)$ has partial derivatives
\[ \partial_t u(x)=\frac{\partial u(t,r)}{\partial t}, \quad
\partial_r u(x)=\frac{\partial u(t,r)}{\partial r},  \]
which are smooth functions with respect to time and coordinate.

Here are three examples of this action.

(a) If  
\[ g_{0} (x,y)=-g_{1} (x,y)=  \delta(x-y) , \]
\be \label{V} V(u(x),u(y))=V(u(x)) \delta(x-y) , \ee
then we get the usual action
\[ S[u]=\int_R d^{2} x \left( 
\frac{1}{2} \left[ \partial_{t} u(x) \right]^2  
- \frac{1}{2} \left[ \partial_{r} u(x) \right]^2 -V(u(x)) \right) . \]

(b) If 
\[ g_{0} (x,y)=- g_1 (x,y)= \delta(x-y)  C_1 (D,r) , \]
\[ V(u(x),u(y))=V(u(x)) \delta(x-y) C_1 (D,r) , \]
where
\[ C_1(D,r)=\frac{|r|^{D-1}}{\Gamma(D)}, \quad (0<D<1) , \]
then we obtain
\[ S[u]=\int dt \int dl_D \, \left( \frac{1}{2} 
\left[ \partial_{t} u(x) \right]^2 - 
\frac{1}{2} \left[ \partial_{r} u(x) \right]^2  -V(u(x)) \right) , \]
where
\[ dl_D=C_1(D,r) d r . \]
This action defines the field $u(x)$ in a medium 
with the fractional Hausdorff dimension $D$ \cite{Tarasov}.

(c) If
\[ V(u(x),u(y))=V(u(x)) \delta(x-y) , \]
\[ g_{0}(x,y)=g_0 \delta(r-r') {\cal K}_0(t,t') , \]
\be \label{g} g_{1}(x,y)=g_1 \delta(t-t') {\cal K}_1(r,r') , \ee
then it follows from (\ref{1}) and (\ref{g}), 
\[ S[u]= 
\frac{1}{2} g_0 \int_{\mathbb{R}} d r  \int_{\mathbb{R}} d t \int_{\mathbb{R}} d t'  \, 
\partial_t u(t,r) {\cal K}_0(t,t')  \partial_{t'} u(t',r) + \]
\be
+\frac{1}{2} g_1 \int_{\mathbb{R}} d t \int_{\mathbb{R}} d r \int_{\mathbb{R}} d r' \,
\partial_{r} u(t,r) {\cal K}_1(r,r')  \partial_{r'} u(t,r') -
\int_{\mathbb{R}} d t \int_{\mathbb{R}} d r \, V(u(t,r)) , \ee
and the time and space dependent kernels are separated in the terms
with derivatives.

We will be interested in a homogeneous case
\[ {\cal K}_1 (r,r')= {\cal K}_1 (r-r') , \]
and an algebraically decaying kernel ${\cal K}_1$ with a power tail, i.e.,
\be \label{G} {\cal K}_1 ( \lambda r)= (\lambda)^{1-\alpha} {\cal K}_1  (r) , 
\quad (1<\alpha<2) . \ee
Similarly, we can consider 
\[ {\cal K}_0 (t,t')={\cal K}_0(t-t')  \]
for $0<t'<t$ as a homogeneous function of order $1-\beta$:
\be \label{T} {\cal K}_0 (\lambda t')=\lambda^{1-\beta} {\cal K}_0 (t') , 
\quad (0< \beta <2, \quad 0<t'<t) . \ee
The relation (\ref{G}) means that we have power-law long-range interaction
in the system. Equation (\ref{T}) indicates the memory effects
with power-law memory function, which can be regarded 
as the influence of the environment.
Just this case of the power-law dependences of ${\cal K}_0 (t)$ and ${\cal K}_1(r)$, 
(\ref{G}) and (\ref{T}), will be considered to derive the 
field equations with fractional derivatives.

\subsection{Gateaux differential and variation of action}

The field equations  will be derived by using
the Gateaux differential \cite{G,F1,Vainberg} of 
$S[u]$ at the point $u(x)$, which is defined as the limit
\be \label{VD0} \delta S[u,h] =
\left( \frac{d}{d \varepsilon} S[u+\varepsilon h] \right)_{\varepsilon=0}=
\lim_{\varepsilon \rightarrow 0} 
\frac{S[u +\varepsilon \ h]-S[u]}{\varepsilon} , \ee
and which exists for fairly smooth integrable  functions $h(x)=\delta u(x)$.
The Gateaux derivative is slightly different 
from the Frechet derivative $\delta_F S[u,h]$, where
\be 
\lim_{\|h\| \rightarrow 0} \frac{\| \, S[u+h]-S[u]- \delta_F S[u,h] \, \| }{\|h\|}=0 .
\ee
The Gateaux derivative is more general concept than Frechet derivative.
If a function is Frechet differentiable, it is also Gateaux differentiable, 
and $\delta S[u,h]$ is a linear operator. 
However, not every Gateaux differentiable function is Frechet differentiable. 
In general, unlike other forms of derivatives, the Gateaux derivative 
is not linear with respect to $h(x)$.

The action (\ref{1}) for $u+\varepsilon h$ is
\[ S[u+\varepsilon h]=\int_R d^{2} x \int_R d^{2} y \Bigl( 
\frac{1}{2} \partial_{t} (u(x) + \varepsilon h(x)) 
g_0 (x,y) \partial_{t'} (u(y) +\varepsilon h(y))+ \]
\[ + \frac{1}{2} \partial_{r} (u(x) + \varepsilon h(x)) 
g_1 (x,y) \partial_{r'} (u(y) +\varepsilon h(y))- 
V(u(x)+\varepsilon h(x),u(y)+\varepsilon h(y)) \Bigr) . \]
This expression up to the order $\varepsilon$ has the form 
\[ S[u+\varepsilon h]=S[u]+\varepsilon \int_R d^{2} x \int_R d^{2} y \Bigl( 
\frac{1}{2} \partial_{t} h(x) g_0 (x,y) \partial_{t'} u(y)+
\frac{1}{2} \partial_{t} u(x) g_0 (x,y) \partial_{t'} h(y)+\]
\[ +\frac{1}{2} \partial_{r} h(x) g_1 (x,y) \partial_{r'} u(y)+
\frac{1}{2} \partial_{r} u(x) g_1 (x,y) \partial_{r'} h(y)- \]
\[ -\frac{\partial V(u(x),u(y))}{\partial u(x)} h(x)-
\frac{\partial V(u(x),u(y)) }{\partial u(y)} h(y) \Bigr) +... \]
In the second, fourth and 6th terms of the right hand side,
we change the variables $x \  \leftrightarrow \ y$. Then
\[ S[u+\varepsilon h]=S[u]+\varepsilon \int_R d^{2} x \int_R d^{2} y \Bigl( 
\frac{1}{2} \partial_{t} h(x) [g_0 (x,y) +g_0 (y,x)] \partial_{t'} u(y)+ \]
\[ +\frac{1}{2} \partial_{r} h(x) [g_1 (x,y) +g_1 (y,x)] \partial_{r'} u(y)- \]
\[ -\frac{\partial [V(u(x),u(y))+V(u(y),u(x))]}{\partial u(x)} h(x) \Bigr) +... \]
It is convenient to introduce the functions
\[ K_{0}(x,y)=\frac{1}{2}[g_0 (x,y) +g_0 (y,x)] , \] 
\be \label{8A}
K_{1}(x,y)=\frac{1}{2}[g_1 (x,y) +g_1 (y,x)] , \ee
\[ U(u(x),u(y))=V(u(x),u(y))+V(u(y),u(x)) . \]
Then the variation of action is
\[ \delta S[u,h] =
\lim_{\varepsilon \rightarrow 0} 
\frac{S[u +\varepsilon \ h]-S[u]}{\varepsilon}=
\int_R d^{2} x \int_R d^{2} y \Bigl( 
\partial_{t} h(x) K_{0} (x,y)  \partial_{t'} u(y)+ \]
\[ +\partial_{r} h(x) K_{1} (x,y)  \partial_{r'} u(y)
-\frac{\partial U(u(x),u(y))}{\partial u(x)} h(x) \Bigr) . \]
Using the relations
\[ \partial_{t} h(x) K_0 (x,y)  \partial_{t'} u(y)=
\partial_{t} \left[ h(x)  K_0 (x,y) \partial_{t'} u(y) \right]-
\partial_{t} \left[ K_0 (x,y)  \partial_{t'} u(y) \right] h(x) , \]
\[ \partial_{r} h(x) K_1 (x,y)  \partial_{r'} u(y)=
\partial_{r} \left[ h(x)  K_1 (x,y) \partial_{r'} u(y) \right]-
\partial_{r} \left[ K_1 (x,y) \partial_{r'} u(y) \right] h(x) , \]
\[ \partial_{t} \partial_{t'} u(y)=\partial_{r} \partial_{r'} u(y)=0  , \]
and the boundary condition 
\[ [h(y)]_{\partial R}=0 , \]
we get
\be \label{osh} \delta S[u,h] =\int_R d^{2} x \, h(x) \int_R d^{2} y  \left(
- \partial_{t} \left[ K_{0} (x,y) \right] \partial_{t'} u(y)
- \partial_{r} \left[ K_{1} (x,y) \right] \partial_{r'} u(y)
-\frac{\partial U(u(x),u(y))}{\partial u(x)} 
\right) . \ee
For the symmetric potential 
\[ U(u(x),u(y))=U(u(x)) \delta(x-y), \]
equation (\ref{osh}) transforms into
\be \label{Z10}
\delta S[u,h] =-\int_R d^{2} x h(x) \left(
\int_R d^{2} y \, \partial_{t} K_{0} (x,y) \, \partial_{t'} u(y)  +
\int_R d^{2} y \, \partial_{r} K_{1} (x,y)  \partial_{r'} u(y)  +
\frac{\partial U(u(x))}{\partial u(x)} \right) . \ee
The dynamical equation follows from the
stationary action principle 
\[ \delta S[u,h]=0 \]
for any $h$. 
The field $u=u(x)$, which leads to a minimum or saddle values of $S[u]$,
describes the space-time evolution. 
For the action (\ref{1}), the stationary principle gives 
\be \label{me} 
\int_R d^{2} y \, \partial_{t} K_{0}(x,y) \, \partial_{t'} u(y) +
\int_R d^{2} y \, \partial_{r} K_{1}(x,y) \, \partial_{r'} u(y) +
\frac{\partial U(u(x))}{\partial u(x)} =0 . \ee
It is an integro-differential equation, 
which allows us to derive field equations for different cases of
the kernels $K_{0}(x,y)$ and $K_1(x,y)$.

\subsection{Special cases}

Let us consider here two special cases: 
(a) system without  memory and with local interaction in space,
(b) field with power-law memory and long-range interaction.

(a) In absence of memory and for local interaction the kernels (\ref{8A})
are defined at the only instant $t$ and point $r$, i.e.,
\[ K_{0}(x,y)=g_{0} \delta(x-y) , \quad  K_{1}(x,y)=g_{1} \delta(x-y) \]
with some constants $g_0$ and $g_1$. 
Then equation (\ref{me}) gives 
\[ g_{0} \partial^2_{t} u(t,r) + g_{1} \partial^2_{r} u(t,r) 
+\frac{\partial U(u(t,r))}{\partial u(t,r)} =0 . \]
For $g_0=1$, $g_1=-1$, and 
\[ U(u(t,r))= -\cos u(t,r) ,  \]
we get the sine-Gordon equation
\be \label{sinG}
\partial^2_t u(t,r)- \partial^2_r u(t,r)+\sin u(t,r)=0  . \ee

(b) In this example, we show how time and space variables
can be separated leaving a possibility to consider the system with 
power-law memory and long-range interaction.
Let $K_0(x,y)$ and $K_1(x,y)$ have the form 
\be \label{MM1} K_0 (x,y)=\delta(r-r') {\cal K}_0 (t,t') , \ee
\be \label{MM2} K_1 (x,y)=\delta(t-t') {\cal K}_1(r,r') , \ee
where $x=(t,r)$, and $y=(t',r')$.  
Then field equation (\ref{me}) can be presented as
\be \label{10} 
Z_t(t,r)+Z_r(t,r)+\frac{\partial U(u(t,r))}{\partial u(t,r)} =0 , \ee
where
\be \label{Zt}
Z_t(t,r)=\int^{+\infty}_{-\infty} dt' \, \partial_{t} {\cal K}_{0} (t,t')\, \partial_{t'} u(t',r) , \ee
\be \label{Zr}
Z_r(t,r)= \int^{+\infty}_{-\infty} dr' \, 
\partial_{r} {\cal K}_{1} (r,r') \frac{\partial u(t,r')}{\partial r'}  \ee
with separated spatial and temporal kernels.
Till now, the kernels ${\cal K}_0(t,t')$ 
and ${\cal K}_1(r,r')$ were not defined.
Their specific choice to present a long-term memory and 
long-range interaction will be in the next two subsections.

\subsection{Power-law memory}

Consider the kernel $\partial_{t} {\cal K}_{0} (t,t')$ of integral (\ref{Zt}) 
in the interval $t' \in (0,t)$ such that
\be \label{Mtt} 
\partial_{t} {\cal K}_{0} (t,t')=
\begin{cases} 
{\cal M} (t-t') , & 0 < t' < t ;
\cr 
0, & t' > t, \quad \ t'< 0 .
\end{cases}
\ee
Then
\be \label{convol}
Z_t(t,r) = \int^{t}_{0} dt' \, {\cal M} (t-t') \, \partial_{t'} u(t',r)=
{\cal M}(t) * \partial_t u(t,r) . \ee
As the result, we have the evolution 
field equation in which the quantity $Z_t(t,r)$ is related 
to another quantity $\partial_{t'} u(t',r)$ through 
a memory function ${\cal M}(t)$.
Equation (\ref{convol}) is a typical non-Markovian equation obtained 
in studying of systems coupled to an environment, 
where environmental degrees of freedom being averaged. 
For a system without memory, we have 
\be \label{Md} {\cal M}(t-t')=\delta(t-t') , \ee
and
\be \label{delta}
Z_t(t,r) = \int^t_0 \delta(t-t') \partial_{t'} u(t',r)  d t' 
=\partial_{t} u(t,r)  , \ee
i.e., the function $Z_t(t,r)$ is defined by $\partial_{t} u(t,r)$ 
at the only current instant $t$. 

Consider now the power-like memory function 
\be \label{Mt1}
{\cal M}(t-t') = \frac{g_0}{\Gamma(1-\beta)} \frac{1}{(t-t')^{\beta}} , 
\quad (0<\beta <1) ,
\ee
where $g_0$ is a constant that can be presented as 
a strength of perturbation induced by the environment,
and $\Gamma(1-\beta)$ is the Gamma function. 

Substitution of (\ref{Mt1}) into (\ref{convol}) gives
\be \label{FracInt}
Z_t(t,r)=\frac{g_0}{\Gamma(1-\beta)} 
\int^t_0 (t-t')^{-\beta} \partial_{t'} u(t',r) dt'=
g_0 \ _{0}^CD^{\beta}_t u (t,r)  , 
\quad (0< \beta<1 ),
\ee
where $\ _{0}^CD^{\beta}_t $ is 
the left fractional Caputo derivative \cite{Podlubny,KST}.

For the kernel $\partial_{t} {\cal K}_{0} (t,t')$ in the integral (\ref{Zt}) such that
\be \label{Mtt2} 
\partial_{t} {\cal K}_{0} (t,t')=
\begin{cases} 
{\cal M}^{\prime} (t'-t) , & t < t' < 0 ;
\cr 
0, & t' > 0, \ \ \ t'< t,
\end{cases}
\ee
where
\be \label{Mt2}
{\cal M}^{\prime}(t'-t) = \frac{(-1)g^{\prime}_0}{\Gamma(1-\beta)} \frac{1}{(t'-t)^{\beta}} ,
\quad (0<\beta<1) ,
\ee
we get
\be 
Z_t(t,r) = \frac{(-1) g^{\prime}_0  }{\Gamma(1-\beta)} 
\int^{0}_{t} \frac{\partial_{t'} u(t',r') }{(t'-t)^{\beta}} dt'=
g^{\prime}_0 \ _{t}^CD^{\beta}_{0}u , \quad (0<\beta<1) ,
\ee
which is the right fractional Caputo derivative \cite{KST,Podlubny}.

In general, the kernel ${\cal K}_0(t,t')$ can include 
positive and negative intervals of time. Then
\be \label{Mtt2pK} 
\partial_t {\cal K}_0(t,t')=
\begin{cases} 
{\cal M} (t-t') , & 0<t'<t  ;
\cr 
{\cal M}^{\prime} (t'-t) , & t<t'< 0 ;
\cr 
0, & 0<t<t', \quad t'<t<0 ,
\end{cases}
\ee
where ${\cal M} (t-t')$ and ${\cal M}^{\prime} (t'-t)$ 
are defined by (\ref{Mt1}) and (\ref{Mt2}).
Then, we get a linear combination of left and right Caputo derivatives
\be \label{ZDD} Z_t(t,r)=
g_0 \ _{0}^CD^{\beta}_{t} u(t,r) + g^{\prime}_0 \ _{t}^CD^{\beta}_{0} u(t,r) , 
\quad (0<\beta<1). 
\ee
As a result, field equation (\ref{me}) consists of fractional time derivatives, 
and it will be written in Sec. 2.6.

We also will be interested in the case when
\be \label{MttK} 
{\cal K}_{0} (t,t')=
\begin{cases} 
{\cal M} (t-t') , & 0 < t' < t ;
\cr 
0, & t' > t, \quad \ t'< 0 ,
\end{cases}
\ee
or
\be \label{Mtt2K} 
{\cal K}_0(t,t')=
\begin{cases} 
{\cal M} (t-t') , & 0<t'<t  ;
\cr 
{\cal M}^{\prime} (t'-t) , & t<t'< 0 ;
\cr 
0, & 0<t<t', \quad t'<t<0 ,
\end{cases}
\ee
(compare to (\ref{Mtt}) and (\ref{Mtt2})),
with the functions ${\cal M}$, ${\cal M}^{\prime}$ 
as in (\ref{Mt1}) and (\ref{Mt2}).
Substitution of (\ref{Mtt}) and (\ref{Mtt2}) into (\ref{Zt}), 
and integration by parts gives, similarly to (\ref{FracInt}) and (\ref{ZDD}),
\be \label{28c} Z_t(t,r)=
g_0 \ _{0}^CD^{\beta+1}_{t} u(t,r)  , 
\quad (0<\beta<1) ,
\ee
or
\be \label{28d} Z_t(t,r)=
g_0 \ _{0}^CD^{\beta+1}_{t} u(t,r) + g^{\prime}_0 \ _{t}^CD^{\beta+1}_{0} u(t,r) , 
\quad (0<\beta<1) 
\ee
with the same field equation (\ref{10}).
Depending on different kernels (\ref{Mtt}), (\ref{Mtt2}) and
(\ref{MttK}), (\ref{Mtt2K}), we obtain field equations with
different order of time derivatives (see in Sec. 2.6).

The Caputo fractional derivatives can be linked to 
fractional powers of variable $s$ for the corresponding 
Laplace-transformed equation.
It is known \cite{C1,C2,Podlubny}, that the Laplace transform 
of the Caputo fractional derivative is
\be \label{33}
\int^{\infty}_0 e^{-s t} \left[ _{0}^CD^{\beta}_t  u(t,r) \right]dt=
s^{\beta} v(s,r) -\sum^{m-1}_{s=0} s^{\beta-q-1} u^{(q)}(0,r) ,
\ee
where $m-1< \beta \le m$, 
\[ u^{(q)}(t,r)=\frac{\partial^q u(t,r)}{\partial t^q} , \]
and $v(s,r)$ is the Laplace transform of $u(t,r)$:  
\be \label{33b}
v(s,r)=\int^{\infty}_0 e^{-s t} u(t,r) dt .
\ee
Note that formula (\ref{33}) involves the initial conditions $u^{(q)}(0,r)$
as integer derivatives $u^{(q)}(t,r)$ with respect to time. 
Therefore we can put the initial conditions in a usual way.
The functions $u(t,r)$ satisfy the condition
\be \label{33c}
\int^{\infty}_{0} e^{-s t} |u(t,r)| < \infty.
\ee
For $0<\beta \le 1$, Eq. (\ref{33}) has the form
\be \label{34}
\int^{\infty}_0 e^{-s t} \left[ _{0}^CD^{\beta}_t  u(t,r) \right] dt=
s^{\beta} v(s,r) -s^{\beta-1} u(0,r) .
\ee
Inversion of (\ref{34}) gives
\be \label{35}
_{0}^CD^{\beta}_t u(t,r)=
\frac{1}{2\pi i} \int_{Br} e^{s t} \left[ s^{\beta} v(s,r) 
-s^{\beta-1} u(0,r) \right] ds ,
\ee
where $Br$ denotes the Bromwich contour.

The final equation of $u(t,r)$ will be written in Sec. 2.6.


\subsection{Non-local interaction}

Consider the kernel ${\cal K}_1 (r,r')$ of the integral (\ref{Zr}) as 
\be \label{Mrr} 
{\cal K}_1 (r,r') = {\cal C} (|r-r'|)=
\frac{-g_1}{\cos (\pi \alpha /2) \Gamma(2-\alpha) }
\frac{1}{|r-r'|^{\alpha-1}} , \quad (1<\alpha<2) 
\ee
that describes the power law interaction. Then, we obtain 
\be 
Z_r(t,r)= \int^{+\infty}_{-\infty} dr' \, {\cal C} (|r-r'|)  
\frac{\partial^2 u(t,r')}{\partial r^{'2}} 
=g_1 \frac{\partial^{\alpha}}{\partial |r|^{\alpha}} u(t,r) ,
\ee
where the fractional Riesz derivative 
with respect to coordinates is introduced \cite{SKM,KST}
(see also Appendix 1).

It is known \cite{SKM} the connection between the 
Riesz fractional derivative and its Fourier transform 
\be
{\cal F}: \quad \frac{\partial^{\alpha}}{\partial |r|^{\alpha}} 
\longrightarrow - |k|^{\alpha} ,
\ee
where ${\cal F}$ is defined by 
\be
\tilde f(k)=({\cal F} \;  f )(k)=\int^{+\infty}_{-\infty} \; f(r) \; e^{-ikr} dr,
\ee
and ${\cal F}^{-1}$ is an inverse Fourier transform 
\be
f(r)=({\cal F}^{-1}  \tilde f )(r) =
\frac{1}{2\pi} \int^{+\infty}_{-\infty} \; \tilde f(k) \; e^{ikr} dk.
\ee
The fractional Riesz derivatives describes properties of fractal media 
or complex media with fractional dispersion law 
(see for example in \cite{Zaslavsky6}).

\subsection{Field equations with fractional derivatives}

Substitution of (\ref{ZDD}) and (\ref{Mrr}) into (\ref{10})
gives the fractional field equation
\be \label{main1}
g_0 \ _{0}^CD^{\beta}_{t} u(t,r) + 
g^{\prime}_0 \ _{t}^CD^{\beta}_{0} u(t,r) + 
g_1 \frac{\partial^{\alpha}}{\partial |r|^{\alpha}} u(t,r)+
\frac{\partial U(u(t,r))}{\partial u(t,r)} =0 ,
\quad (1<\alpha<2, \ 0<\beta<1) .
\ee
This equation describes the field of the system with power-law memory
and long-range interaction. 
Depending on the situation, $g_0$ or $g^{\prime}_0$ could be zero or not. 
For example, the potential
\[ U(u(t,r))=\frac{a}{2} u^2(t,r)+\frac{b}{4} u^4(t,r)  \]
in Eq. (\ref{main1}) gives the fractional time-dependent 
generalization of the Ginzburg-Landau equation
\be 
g_0 \ _{0}^CD^{\beta}_t u(t,r) + 
g^{\prime}_0 \ _{t}^CD^{\beta}_{0} u(t,r) +
g_1 \frac{\partial^{\alpha}}{\partial |r|^{\alpha}} u(t,r)+
a u(t,r)+b u^3(t,r) =0 , \quad (1<\alpha<2, \ 0<\beta<1) . 
\ee

In the case of the time kernel (\ref{Mtt2K}), Eq. (\ref{main1})
is replaced by
\be \label{44a}
g_0 \ _{0}^CD^{\beta+1}_{t} u(t,r) + 
g^{\prime}_0 \ _{t}^CD^{\beta+1}_{0} u(t,r) + 
g_1 \frac{\partial^{\alpha}}{\partial |r|^{\alpha}} u(t,r)+
\frac{\partial U(u(t,r))}{\partial u(t,r)} =0 ,
\quad (1<\alpha<2, \ 0<\beta<1) .
\ee
This equation has increased by one the order of 
time derivative and can be applied to the wave propagation in media with 
fractional dispersion law.
Particularly in the case when only right time derivative
should be used (one-directional wave propagation)
and the potential is
\[ U(u(t,r))=-\cos  u (t,r) , \]
Eq. (\ref{44a}) gives the fractional sine-Gordon equation
\be 
\ _{0}^CD^{\beta+1}_t u(t,r) - 
\frac{\partial^{\alpha}}{\partial |r|^{\alpha}} u(t,r)+ \sin u(t,r) =0 ,
\quad (1<\alpha<2, \ 0<\beta<1) ,
\ee
where we put $g_0=1$, $g^{\prime}_0=0$ and $g_1=-1$ , and
which is a generalization of (\ref{sinG}) for noninteger derivatives 
with respect to time and coordinate.

Finally, let us simplify the notation and write down 
Eqs. (\ref{main1}) or (\ref{44a}) as
\be
g \frac{\partial^{\beta} }{\partial t^{\beta}}u(t,r)+g_1
\frac{\partial^{\alpha}}{\partial |r|^{\alpha}} u(t,r)+ U^{\prime}(u(t,r))=0 ,
\ee
where $\partial^{\beta}/\partial t^{\beta}$ stays for left, right, or
both Caputo derivatives 
(in the latter case, the constant $g$ 
can be different for different derivatives), and
$0<\beta<2$, $0<\alpha<2$, and $U^{\prime}(u)=\partial U/\partial u$.
Let us comment that the choice of the derivative 
$\partial^{\beta}/\partial t^{\beta}$ depends on 
the type of initial conditions and the processes,
and other than Caputo derivative can appear.

In Appendix 2, we present a generalization of Eqs. ({43}), (\ref{44a}) 
for the $n$-dimensional coordinate case.


\section{Fractional Ginzburg-Landau equation}

Since the variable $x$ in (\ref{1}) can be not specified,
one can apply a similar technique to other problems,
defined by the extremum of a functional with long-range interaction.
As an example, consider a free energy functional for a model of
Ginzburg-Landau equation (GLE) that consists of
long-range interaction.

The fractional generalization of the Ginzburg-Landau equation (FGLE)
was suggested in Ref. \cite{Zaslavsky6}.
This equation can be used to describe the processes 
in complex media \cite{TZ1,Mil}. 
Some properties of FGLE are discussed in \cite{TZ3,Psi,TZ2}. 

It is known \cite{LP}  that the stationary GLE 
\[ g \Delta u-au-bu^3=0 \]
can be derived as the variational Euler-Lagrange equation
\be \label{FGL1a}  \frac{\delta F[u]}{\delta u (r)}=0  \ee
for the free energy functional
\be \label{FGL2a} F[u]=F_0+\frac{1}{2}\int_R 
[g( \partial u)^2 + au^2+\frac{b}{2}u^4 ] d r , \ee
where $\partial u =\partial u(r) / \partial r$, and 
the integration is over a region $R$. 
Here $F_0$ is a free energy of the normal state, 
i.e. $F[u]$ for $u=0$.

Consider the thermodynamic potential (free energy
functional) $F[u]$ for the non-equilibrium state of
a medium with power-law non-local interaction. 
The generalized free energy functional has the form
\be \label{F} F[u]=F_0+\int_R d r \int_R d r' \,
{\cal F}(u(r),u(r'),\partial u(r), \partial u(r'))  ,\ee
where the generalized density of free energy
\[ {\cal F}(u(r),u(r'),\partial u(r), \partial u(r'))  = \]
\be \label{F2a} 
=\frac{1}{2} g {\cal K}_1(r,r') \partial^{k}_{r} u(r) \partial^{k}_{r'} u(r') +
\left( \frac{a}{2} u^2(r)+\frac{b}{4}u^4(r) \right) \delta(r-r')  \ee
has the kernel ${\cal K}_1(r,r')$ defined as in (\ref{Mrr}).
The variational equation (\ref{FGL1a}) gives 
\be 
g \frac{\partial^{\alpha}}{\partial |r|^{\alpha}} u(r)+a u(r)+b u^3(r) =0 ,\quad 
(1<\alpha<2) , 
\ee
which can be called the $\alpha$-FGLE.
This equation can be easily generalized for the $3$-dimensional variable
${\bf r}$ (see Appendix 2).

In the non-stationary case, Eq. (\ref{FGL1a}) should be replaced by
\be \label{88a}  \frac{\partial u(t,{\bf r})}{\partial t}=
\frac{\delta F[u]}{\delta u (t,{\bf r})} , \ee
(see \cite{Tabor}), where there is no explicit time memory effects.
To put such memory into (\ref{88a}), we can write
\be
\int^{t}_{0} dt' {\cal M} (t-t') \frac{\partial u(t',{\bf r})}{\partial t'}=
\frac{\delta F[u(t,{\bf r})]}{\delta u (t,{\bf r})} , 
\ee
and to assume for ${\cal M}(t-t')$ the power law (\ref{Mt1}).
Then we arrive to a nonstationary generalization of $(\alpha,\beta)$-FGLE
\be \label{88c}
\frac{\partial^{\beta} u(t,{\bf r})}{\partial t^{\beta}}=
g \frac{\partial^{\alpha}}{\partial |{\bf r}|^{\alpha}} u(t,{\bf r})
+a u(t,{\bf r})+b u^3(t,{\bf r}) , \quad 
(0<\beta<1, \quad 1<\alpha <2),
\ee
where $\partial^{\beta} / \partial t^{\beta}$ 
is used for Caputo derivative while any other
fractional derivative can be applied by modifying the memory kernel
${\cal M}(t)$, and initial conditions.

It is worthwhile to compare Eq. (\ref{88c}) 
to its counterpart nonlinear Schr$\ddot{o}$dinger equation (NSE)
\be \label{88d}
i\frac{\partial u}{\partial t} = g\Delta u+au +b|u|^2 u
\ee
with $u=u(t,{\bf r})$, and complex $a$, $b$.
In the case of $\Delta_{\perp}$ instead of $\Delta$, Eq. (\ref{88d})
also known as parabolic equation for wave propagation.
Generalization of (\ref{88d}) for the case of fractional space derivative
and non-local interaction ($\alpha$-NLS) was considered in \cite{Zaslavsky6,TZ2}:
\be \label{88e}
i\frac{\partial u}{\partial t} =- g(-\Delta)^{\alpha/2} u+au +b|u|^2 u,
\quad (1< \alpha <2) ,
\ee
where the fractional Laplacian is defined through 
the Fourier transform and Riesz derivatives \cite{SKM}:
\be \label{88f}
{\cal F}: \quad (-\Delta )^{\alpha /2} 
\longrightarrow
({\bf k}^2 )^{\alpha /2} .
\ee
Similar to (\ref{88c}) generalized $(\alpha,\beta)$-NLS
equation has the form
\be \label{89a}
\frac{\partial^{\beta} u}{\partial t^{\beta}} =
- g(-\Delta)^{\alpha/2} u+au +b|u|^2 u,
\ee
where $\partial^{\beta}/\partial t^{\beta}$ 
is now Riemann-Liouville derivative with the Fourier transform
\be \label{89b}
{\cal F}: \quad  \frac{\partial^{\beta} }{\partial t^{\beta}} 
\longrightarrow (i \omega )^{\beta} ,
\ee
and the memory function is working through the parameter $\beta$.

It is convenient also to interpret (\ref{89a}) through the nonlinear
dispersion law by applying to (\ref{89a}) Fourier transform
in both time and space. Then it gives with the help of (\ref{88f})
and (\ref{89b})
\be \label{89c}
(i\omega)^{\beta}=-g({\bf k}^2)^{\alpha/2}+a+b|u| ,
\ee
which was derived for $\beta=1$ in \cite{Zaslavsky6,TZ2}.
Onset of fractional time derivative in (\ref{89a}) 
can stop self-focusing of waves, steepening of the solution,
developing of a singularity.
These phenomena need a special analysis.

Eq. (\ref{89a}) can be easily generalized for the anisotropic case
\be
\frac{\partial^{\beta} u}{\partial t^{\beta}} =
- g_{\perp}(-\Delta_{\perp})^{\alpha_{\perp}/2} u-
g_{\parallel} (-\Delta_{\parallel} )^{\alpha_{\parallel} /2} u+au +b|u|^2 u
\ee
with a corresponding anisotropic dispersion equation instead of (\ref{89c})
(see also \cite{Zaslavsky6,TZ2} for $\beta=1$).


\section{Discrete system with memory and long-range interaction}

\subsection{Equation for discrete chains}

In this section, we show how the obtained results of Sec. 2 
can be applied to discrete systems, for example chains of
interacting particles.

Long-range interaction is a subject 
of a great interest since a long time. 
Thermodynamics of a model of classical spins with long-range
interactions has been considered in \cite{Dyson,J,CMP,NakTak,S}.
The long-range interactions have been 
widely studied in discrete systems of lattices 
as well as in their continuous analogues:
solitons in one-dimensional lattice with the 
Lennard-Jones-type interaction \cite{Ish};
kinks in the Frenkel-Kontorova model \cite{BKZ};
time periodic spatially localized solutions (breathers) \cite{Br4,GF}; 
energy and decay properties of discrete breathers 
in the framework of the Klein-Gordon equation \cite{BK}, 
and discrete nonlinear Schr$\ddot{o}$dinger equations \cite{Br6}. 
A remarkable property of the dynamics described by the equation with 
fractional space derivatives is that the 
solutions have power-like tails. 
Similar features were observed 
in the lattice models with power-like long-range 
interactions \cite{PV,Br4,GF,AEL,AK,KZT}. 
Long-range interaction can be relevant to the systems such 
as neuron populations \cite{DJD} and Josephson junctions \cite{WCS}.
The syncronization of chaotic systems with power-law long-range 
interactions were considered in \cite{WCS,TZ3,TCT}.
A model of coupled map lattices with coupling that decays in 
a power-law, was considered in \cite{TL,APBV,AT,TCT}.
Note that the fractional power-law dependence can be linked to 
fractal properties of heterogeneous surfaces \cite{PA83}
and power-law decay of structure factor for 
geometry of colloids aggregates \cite{SMWC}.
It will be shown how long-range coupling of particles
and memory function with power tails can reveal a new type of particle
equations with fractional derivatives and the connection 
of these equations to their continuous media counterpart.

Consider an one-dimensional chain of interacting oscillators 
that can be described by the action,
\be \label{Main_Eq}
S[u]= \int^{+\infty}_{-\infty} dt \int^{+\infty}_{-\infty} dt' 
\sum^{+\infty}_{n=-\infty} {\cal L}(u_n(t),u_n(t'),\dot{u}_n(t),\dot{u}_n(t')) ,
\ee
where $u_n$ are displacements of the oscillators from the equilibrium and
${\cal L}$ is a Lagrangian. 
If 
\[ {\cal L}(u_n(t),u_n(t'),\dot{u}_n(t),\dot{u}_n(t'))=
{\cal L}(u_n(t),\dot{u}_n(t)) \, \delta(t-t') , \]
then we have the chain without memory.

Let us introduce a generalization of (\ref{Main_Eq}) with the action
\[ S[u_n]=\int^{+\infty}_{-\infty} dt \int^{+\infty}_{-\infty} dt' \Bigl(
\sum_{n=-\infty}^{+\infty}
\left[ \frac{1}{2} \; {\cal K}_0(t,t') \dot{u}_n(t) \dot{u}_n(t')
-V (u_n(t),u_n(t')) \right] - \]
\be \label{Sun}
-\sum_{\substack{n, m=-\infty \\ m \ne n}}^{+\infty} \; 
U (u_n(t),u_m(t')) \Bigr) .
\ee
In the same way as in Sec. 2, let us separate the kinetic energy from 
the long-range interaction and potential parts:
\be \label{UU}
U(u_n(t),u_m(t'))= \frac{1}{4} g_0 J_{\alpha}(|n-m|) \, (u_n(t)-u_m(t))^2 \, \delta(t-t') ,
\ee
\be \label{VVV}
V(u_n(t),u_n(t'))=V(u_n(t)) \, \delta(t-t') .
\ee
Note that (\ref{VVV}), (\ref{UU}) in (\ref{Main_Eq}) are equivalent to
\be \label{UU2}
U(u_n(t),u_m(t'))=- \frac{1}{2} g_0 J_{\alpha}(|n-m|) \, u_n(t)u_m(t) \, \delta(t-t') ,
\ee
\be \label{VVV2}
V(u_n(t),u_n(t'))=\left( V(u_n(t))+ \frac{1}{2} \tilde g u^2_n(t) \right) \, \delta(t-t') ,
\ee
where
\[ \tilde g= g_0 \sum_{m\not=0} J_{\alpha}(|m|)  . \]
The second term in the right hand side of (\ref{VVV2})
removes the infinity of the interaction (\ref{UU2}) in the continuous medium limit.
The interparticle interaction $J_{\alpha}(|n-m|)$ in (\ref{UU}) 
is defined by
\be \label{189}
J_{\alpha}(|n-m|) =\frac{1}{|n-m|^{\alpha+1}}, \quad (\alpha>0) .
\ee
Some other examples of functions $J_{\alpha}(n)$ can be found in \cite{JMP}.


Using (\ref{Mtt}) and (\ref{Mt1}), for 
the kernel ${\cal K}_0(t,t')$ in (\ref{Sun}), i.e.
\be 
\partial_{t} {\cal K}_{0} (t,t')=
\begin{cases} 
\frac{g_0}{\Gamma(1-\beta)} (t-t')^{-\beta} ,
& 0 < t' < t \quad (0<\beta <1) ;
\cr 
0, & t' > t, \quad \ t'< 0 ,
\end{cases}
\ee
we obtain the corresponding Euler-Lagrange equations
\be \label{Z2a}
 _{0}^CD^{\beta}_{t} u_n(t)+
g_0 \sum_{\substack{m=-\infty \\ m \ne n}}^{+\infty} \; 
J_{\alpha}(|n-m|) \; [u_m(t)-u_n(t)] + \; F (u_n(t))=0 ,
\ee
where $F(u)=\partial V (u)/ \partial u$.

A continuous limit of equation (\ref{Z2a}) can be defined by
a transform operation from $u_n(t)$ to $u(x,t)$
\cite{LZ,TZ3,KZT,KZ,JMP}. 
First, define $u_n(t)$ as Fourier coefficients of some 
function $\hat{u}(k,t)$, $k \in [-K/2, K/2]$, i.e.
\be \label{ukt}
\hat{u}(t,k) = \sum_{n=-\infty}^{+\infty} \; u_n(t) \; e^{-i k x_n} =
{\cal F}_{\Delta} \{u_n(t)\} ,
\ee
where  $x_n = n \Delta x$, and $\Delta x=2\pi/K$ is a
distance between nearest particles in the chain, and
\be \label{un} 
u_n(t) = \frac{1}{K} \int_{-K/2}^{+K/2} dk \ \hat{u}(t,k) \; e^{i k x_n}= 
{\cal F}^{-1}_{\Delta} \{ \hat{u}(t,k) \} . 
\ee
Secondly, in the limit $\Delta x \rightarrow 0$ ($K \rightarrow \infty$) 
replace $u_n(t)=(2\pi/K) u(x_n,t) \rightarrow  u(x,t) dx$, 
and  $x_n=n\Delta x= 2\pi n/K \rightarrow x$.
In this limit, Eqs. (\ref{ukt}), (\ref{un}) are transformed into 
the integrals
\be \label{ukt2} 
\tilde{u}(t,k)=\int^{+\infty}_{-\infty} dx \ e^{-ikx} u(t,x) = 
{\cal F} \{ u(t,x) \} = \lim_{\Delta x \rightarrow 0} {\cal F}_{\Delta} \{u_n(t)\} , 
\ee
\be \label{uxt}
u(t,x)=\frac{1}{2\pi} \int^{+\infty}_{-\infty} dk \ e^{ikx} \tilde{u}(t,k) =
 {\cal F}^{-1} \{ \tilde{u}(t,k) \}= 
\lim_{\Delta x \rightarrow 0} {\cal F}^{-1}_{\Delta} \{ \hat{u}(t,k) \}  . 
\ee

Applying (\ref{ukt}) to (\ref{Z2a})
and performing the limit (\ref{ukt2}), we obtain
\be \label{200}
\frac{\partial^{\beta} u(t,x)}{\partial t^{\beta}}+
g_{\alpha} \frac{\partial^{\alpha} u(t,x)}{\partial |x|^{\alpha}}+
F(u(t,x))=0, \quad (0<\beta<2, \ 1<\alpha<2) ,
\ee
where 
\be
g_{\alpha}=2 g_0 (\Delta x)^{\alpha}  \Gamma(-\alpha) \cos 
\left( \frac{\pi \alpha}{2} \right)
\ee
is the renormalized constant. 
The Caputo time derivative is written in a simplified form 
$\partial^{\beta} / \partial t^{\beta}$, 
and the value of $\beta$ depends on the choice of memory function. 
The equation (\ref{200}) can be generalized to a nonlinear long-range interaction.
Consider, instead of (\ref{Z2a}),
\be \label{211}
 _{0}^CD^{\beta}_{t} u_n+
g_0 \sum_{\substack{m=-\infty \\ m \ne n}}^{+\infty} \; 
J_{\alpha}(|n-m|) \; [f(u_m)-f(u_n)] + \; F (u_n)=0 ,
\ee
where $f(u)$ is a function of $u$. 
For example, $f(u)=u^2$ or $f(u)=u-gu^2$.
Then the corresponding continuous limit for the same $J_{\alpha}(|n-m|)$
as (\ref{189}) leads to the time-space fractional equation
\be \label{212}
\frac{\partial^{\beta} u(t,x)}{\partial t^{\beta}}+
g_{\alpha} \frac{\partial^{\alpha}}{\partial |x|^{\alpha}} f(u(t,x))+
F(u(t,x))=0, \quad (0<\beta<2, \ 1<\alpha<2) .
\ee
Equations (\ref{200}) and (\ref{212}) for $\beta=1$ were considered in 
\cite{LZ,TZ3,KZT,KZ,JMP}. 
Generalization to $0<\beta <2$ significantly extends the area of their
applications.
A physical motivation is that a dynamical process typically reveals 
fractional features simultaneously in space and time.
Such situation just was considered in chaotic dynamics \cite{Zaslavsky7,Zaslavsky1}.
Now we have such a possibility far beyond the fractional kinetics.
An evident generalization of (\ref{212}) is for the inter-particle interactions
with two or more different kernels.
For example one can consider regular terms without long memory
together with a term with long memory:
\be \label{212b}
\frac{\partial^{\beta} u(t,x)}{\partial t^{\beta}}+
g_{s} \frac{\partial^{s} u(t,x)}{\partial |x|^{s}}+
g_{\alpha} \frac{\partial^{\alpha} u(t,x)}{\partial |x|^{\alpha}} +
F(u(t,x))=0, \quad (0<\beta<2, \ 1<\alpha<2) 
\ee
with some $g_s$ and $g_{\alpha}$ and integer $s$.

\section{Conclusion}

Starting from a variation of the action functional, 
we consider different type of kernels that define the
character of particle interaction and
the influence of an environment on the memory function.
The main stress is on the long-range interaction and memory
that occur in complex media.
The case when the interaction or memory function have power-law
structure the system can be described by the equation of motion 
with fractional derivatives $\partial^{\beta}/\partial t^{\beta}$
and $\partial^{\alpha}/\partial |x|^{\alpha}$
depending on the power of interaction and  memory function.
We have discussed how different types of
the derivatives and possible values $(\alpha,\beta)$ 
may occur with respect to the type of memory and interaction.
The final equations of motions can be considered as a new kind
of tool to study dynamics with space-time distributed interactions.
Number of examples of such kind of systems can be found in the reviews
\cite{Montr,Zaslavsky1} related to random or chaotic processes.
The study of this paper shows that the list of possible
applications of fractional equations can be naturally expanded
to include non-chaotic and non-random dynamics as well.

This work was supported by the Office of Naval Research, 
Grant No. N00014-02-1-0056, and the NSF Grant No. DMS-0417800.


\newpage


\section*{Appendix 1: Fractional derivatives}

The fractional derivative has different definitions \cite{SKM,OS}, 
and exploiting any of them depends on the kind of 
the problems, initial (boundary) conditions, and 
the specifics of the considered physical processes.
The classical definition is the so-called Riemann-Liouville
derivative \cite{SKM,OS}.
The left and right Riemann-Liouville derivatives for an 
interval $[a,b]$ are defined by
\[
_a{\cal D}^{\alpha}_{t}u(x)=
\frac{1}{\Gamma(n-\alpha)} \frac{\partial^n}{\partial x^n}
\int^{x}_{a} \frac{u(z) dz}{(x-z)^{\alpha-n+1}},
\]
\be \label{P1}
_t{\cal D}^{\alpha}_{b}u(x)=
\frac{(-1)^n}{\Gamma(n-\alpha)} \frac{\partial^n}{\partial x^n}
\int^{b}_x \frac{u(z) dz}{(z-x)^{\alpha-n+1}} ,
\ee
where $n-1<\alpha<n$.
End-points $a$, $b$ can be extended to $-\infty$, $\infty$
if the integral exists.

Due to reasons, concerning the initial conditions,
it is more convenient to use the Caputo fractional derivatives
\cite{Podlubny}.
Its main advantage is that the initial conditions take the same
form as for integer-order differential equations.
The Caputo fractional derivatives are 
\[
_a^CD^{\alpha}_{x}u(x)=
\frac{1}{\Gamma(n-\alpha)} 
\int^{x}_{a} \frac{u^{(n)}(z) dz}{(x-z)^{\alpha-n+1}},
\]
\be \label{P1c}
_x^CD^{\alpha}_{b}u(x)=
\frac{(-1)^n}{\Gamma(n-\alpha)} 
\int^{b}_x \frac{u^{(n)}(z) dz}{(z-x)^{\alpha-n+1}} ,
\ee
where $u^{(n)}(z)=d^n u(z)/dz^n$, and $n-1<\alpha<n$.
The Caputo fractional derivatives can be 
defined through the Riemann-Liouville derivatives  \cite{KST} by 
\[
_a^CD^{\alpha}_xu(x)=\ _a{\cal D}^{\alpha}_x \left( u(x)-
\sum^{n-1}_{k=0} \frac{(x-a)^k}{k!} u^{(k)}(a)  \right) ,
\]
\be \label{P2}
_x^CD^{\alpha}_b u(x)=\ _x{\cal D}^{\alpha}_b \left( u(x)-
\sum^{n-1}_{k=0} \frac{(b-x)^k}{k!} u^{(k)}(b)  \right) ,
\ee
where $n-1<\alpha<n$. These equations give
\be \label{RL2}
_a{\cal D}^{\alpha}_x u(x)= \ _a^CD^{\alpha}_x u(x)+
\sum^{n-1}_{k=0} \frac{(x-a)^{k-\alpha}}{\Gamma(k-\alpha+1)} u^{(k)}(a) ,
\ee
\be \label{RL2}
_x{\cal D}^{\alpha}_b u(x)= \ _x^CD^{\alpha}_b u(x)+
\sum^{n-1}_{k=0} \frac{(b-x)^{k-\alpha}}{\Gamma(k-\alpha+1)} u^{(k)}(b) ,
\ee

The Riesz fractional derivative of order $\alpha$ are
\be \label{P4}
\frac{\partial^{\alpha}}{\partial |x|^{\alpha}} u(x)=
-\frac{1}{2 \cos(\pi \alpha /2)} 
\left({\cal D}^{\alpha}_{+}u(x) +{\cal D}^{\alpha}_{-}u(x)\right) ,
\ee
where $\alpha\not=1,3,5...$, and ${\cal D}^{\alpha}_{\pm}$ are 
Riemann-Liouville fractional derivatives with infinite limits:
\[
{\cal D}^{\alpha}_{+}u(x)=
\frac{1}{\Gamma(n-\alpha)} \frac{\partial^n}{\partial x^n}
\int^{x}_{-\infty} \frac{u(z) dz}{(x-z)^{\alpha-n+1}},
\]
\be \label{P5}
{\cal D}^{\alpha}_{-}u(x)=
\frac{(-1)^n}{\Gamma(n-\alpha)} \frac{\partial^n}{\partial x^n}
\int^{\infty}_x \frac{u(z) dz}{(z-x)^{\alpha-n+1}} .
\ee
Substitution of Eqs. (\ref{P5}) into Eq. (\ref{P4}) gives
\be \label{P6}
\frac{\partial^{\alpha}}{\partial |x|^{\alpha}} u(x)=
\frac{-1}{2 \cos(\pi \alpha /2) \Gamma(n-\alpha)} 
\frac{\partial^n}{\partial x^n} 
\left(
\int^{x}_{-\infty} \frac{u(z) dz}{(x-z)^{\alpha-n+1}}+
\int^{+\infty}_x \frac{(-1)^n u(z) dz}{(z-x)^{\alpha-n+1}}
\right) .
\ee

The Fourier transform of the fractional derivatives \cite{SKM,KST} are
\be
{\cal F} \left( \frac{\partial^{\alpha}}{\partial |x|^{\alpha}} u(x) \right)(k) 
= - |k|^{\alpha} \tilde u(k),
\ee
\be
{\cal F} \left( {\cal D}^{\alpha}_{\pm} u(x) \right)(k) 
= (\pm i k )^{\alpha} \tilde u(k),
\ee
where ${\cal F}$ is defined by 
\be
\tilde u(k)=({\cal F} \; u )(k)=\int^{+\infty}_{-\infty} \; u(x) \; e^{-ikx} dx .
\ee
The inverse Fourier transform is 
\be
u(x)=({\cal F}^{-1}  \tilde u )(x) =
\frac{1}{2\pi} \int^{+\infty}_{-\infty} \; \tilde u(k) \; e^{ikx} dk.
\ee

\section*{Appendix 2: n-dimensional case}

The generalization the action (\ref{1}) for the case $r \in \mathbb{R}^n$, 
where $x=(t,r)$, and $r=(x^1,...,x^n)$, gives the field equation
\be  
g_0 \ _{0}^CD^{\beta}_{t} u(t,r) + 
g^{\prime}_0 \ _{t}^CD^{\beta}_{0} u(t,r) + \sum^n_{k=1}
g_k \frac{\partial^{\alpha}}{\partial |x^k|^{\alpha_k}} u(t,r)+
\frac{\partial U(u(t,r))}{\partial u(t,r)} =0 . 
\ee

For the case $r \in \mathbb{R}^n$, there exists the other possibility 
to define the kernels $M_t(x,y)$ and $M_r(x,y)=\{M_k(x,y), k=1,...,n \}$. 
We can consider  
\be  \label{Zrn}
Z_r(t,r)= \int_{\mathbb{R}^n} d^nr' \, 
\sum^n_{k=1} M_k (r,r') \frac{\partial u(t,r')}{\partial x^{\prime k}} ,
\ee
where $M_k (r,r')$ is Riesz kernel \cite{SKM}:
\[
M_k (r,r')= K_{\alpha_k}(r-r')=\frac{1}{\gamma_n(\alpha_k)} \ 
\begin{cases}
|r-r'|^{\alpha_k-n} & \alpha_k-n \not=0,2,4,...
\cr
-|r-r'|^{\alpha_k-n} \; \ln |r-r'| & \alpha_k-n =0,2,4,...
\end{cases}
\]
Here $\alpha_k>0$, ($\alpha_k \not=n,n+2,n+4,...$), and
\be
\gamma_n(\alpha)=
\begin{cases}
2^{\alpha} \pi^{n/2}\Gamma(\alpha/2)/ \Gamma(\frac{n-\alpha}{2}) &
\alpha \not=n+2k, \quad n \not= -2k,
\cr
1 & n = -2k,
\cr
(-1)^{(n-\alpha)/2}2^{\alpha-1} \pi^{n/2} \;
\Gamma(\alpha/2) \; 
\left[ \frac{\alpha-n}{2} \right]! 
 & \alpha \not=n+2k.
\end{cases}
\ee
Note that the multivariable Riesz integral 
\be
(I^{\alpha} u)(t,r)=\frac{1}{\gamma_n(\alpha)}
\int_{\mathbb{R}^n} \frac{u(t,r') dr'}{|r-r'|^{n-\alpha}},
\ee
where $\alpha>0$, can be presented as convolution:
\be
(I^{\alpha} u)(t,r) =\int_{\mathbb{R}^n} \; K_{\alpha}(r-r')\; u(t,r') \; d^n r',
\ee
with the Riesz kernel $K_{\alpha}(r)$. 
It allows us to write (\ref{Zrn}) as
\be Z_r(t,r)= \sum^n_{k=1} I^{\alpha_k} \frac{\partial u(t,r)}{\partial x^k}  . \ee
The fractional Riesz integrals of orders $\alpha_k$ ($k=1,...,n$) 
in the field equations describe the fractal media.  
Then the field equation is
\be 
g_0 \ _{0}^CD^{\beta}_{t} u(t,r) + 
g^{\prime}_0 \ _{t}^CD^{\beta}_{0} u(t,r) + 
\sum^n_{k=1} I^{\alpha_k} \frac{\partial u(t,r)}{\partial x^k} +
\frac{\partial U(u(t,r))}{\partial u(t,r)} =0 . 
\ee

If $M_k(r,r')$ in (\ref{Zrn}) is an operator such that
\be 
M_k(r,r') u(t,r')=g_k \frac{1}{d_{n,l}(\alpha_k)} 
\frac{(\Delta^l_{r'}u)(t,r) }{|r'|^{n+\alpha_k}} ,
\ee
where 
\[ d_{n,l}(\alpha)=  
\frac{2^{-\alpha} \pi^{1+n/2} }{ \Gamma(1+\alpha/2) \Gamma((n+\alpha)/2) \sin(\alpha \pi/2)} 
\sum^l_{k=0} \frac{(-1)^{k-1} l !}{ (l-k)! k!} \, k^{\alpha} \]
is normalized multiplier \cite{KST}, and
\[ (\Delta^l_{r'} u)(t,r)=\sum^l_{k=0} (-1)^{k-1} \frac{l!}{(l-k)! k!} u(t,r-kr') \]
is symmetrized difference \cite{KST}, then
\[ Z_r(t,r)= \sum^n_{k=1} g_k
\frac{\partial^{\alpha_k}}{\partial |r|^{\alpha_k}} \frac{\partial u(t,r)}{\partial x^k} . \]
As a result, we have 
\be \label{main2}
g_0 \ _{0}^CD^{\beta}_{t} u(t,r) + 
g^{\prime}_0 \ _{t}^CD^{\beta}_{0} u(t,r) + 
g_k \sum^n_{k=1} \frac{\partial^{\alpha_k}}{\partial |r|^{\alpha_k}} 
\frac{\partial u(t,r)}{\partial x^k}  +
\frac{\partial U(u(t,r))}{\partial u(t,r)} =0 ,
\ee
which is the field equations 
with $n$ fractional Riesz multivariable derivatives.

\end{document}